\documentclass[aoas,nameyear,dvips]{arximspdf}
\usepackage{multirow}
\usepackage{graphicx}

\doi{10.1214/09-AOAS293}
\volume{4}
\issue{2}
\pubyear{2010}
\firstpage{589}
\lastpage{614}

\makeatletter

\newtheorem{theorem}{Theorem}
\def\eqref#1{(\ref{#1})}
\makeatother

\begin{document}
\begin{frontmatter}

\title{Empirical stationary correlations for semi-supervised learning on graphs\protect\thanksref[1]{T1}}
\thankstext{T1}{Supported by NSF Grant DMS-06-04939.}
\runtitle{Empirical stationary correlations on graphs}
\pdftitle{Empirical stationary correlations
for semi-supervised learning on graphs}

\begin{aug}
\author[A]{\fnms{Ya} \snm{Xu}\thanksref[2]{t2}},
\author[B]{\fnms{Justin S.} \snm{Dyer}\thanksref[3]{t3}}
\and
\author[C]{\fnms{Art B.} \snm{Owen}\ead[label=e3]{owen@stat.stanford.edu}\corref{}}
\pdfauthor{Ya Xu, Justin S. Dyer, Art B. Owen}
\thankstext{t2}{Supported by a Stanford Graduate Fellowship
and the Yahoo! Key Scientific Challenges program.}
\thankstext{t3}{Supported by a National
Science Foundation Graduate Research Fellowship.}
\runauthor{Y. Xu, J. S. Dyer and A. B. Owen}
\affiliation{Stanford University}
\address[C]{Department of Statistics\\
Stanford University\\ Sequoia Hall\\ Stanford, California 94305\\
USA\\ \printead{e3}} 
\end{aug}

\received{\smonth{1} \syear{2009}}
\revised{\smonth{7} \syear{2009}}

%
\begin{abstract}
In semi-supervised learning on graphs, response variables observed
at one node are used to estimate missing values at
other nodes. The methods exploit correlations
between nearby nodes in the graph.
In this paper we prove that many such proposals are
equivalent to kriging predictors based on a fixed covariance
matrix driven by the link structure of the graph.
We then propose a data-driven estimator of the
correlation structure that exploits patterns among the
observed response values.
By incorporating even a small fraction of observed
covariation into the predictions, we are able to obtain
much improved prediction on two graph data sets.
\end{abstract}

%

\end{frontmatter}

\section{Introduction}\label{sec1}
Data on graphs has long been with us, but the recent explosion
of interest in social network data available
on the Internet has brought this sort of data to
prominence.
A typical problem is to predict the value of a feature
at one or more nodes in the graph. That feature is assumed to
have been measured on some, but not all, nodes of the graph.
For example, we might want to predict which web pages
are spam, after a human expert has labeled a subset
of them as spam or not.
Similarly, we might want to know on which Facebook
web pages an ad would get a click, although that
ad has only been shown on a subset of pages.

The underlying assumption in these prediction problems is
that there is some correlation, usually positive, between
the values at vertices that are close to each other in the graph.
By making predictions that are smooth with respect to a
notion of distance in the graph, one is able to define
a local average prediction.

This problem is often called semi-supervised learning,
because some of the data have measured response
values, while others have predictor only.
We suppose that the response random variable at node $i$ of the
graph is $Y_i$.
The observed value $y_i$ is available at
some, but not all, $i$.
For a survey of semi-supervised learning see \citet{Zhu2005}.

Our starting point is the graph based random walk
strategy of \citet{Zhou2005a}.
To describe their approach,
let $G$ be a weighted directed graph with $n$ vertices.
The edges of $G$ are represented by an adjacency
matrix $W$ with entries $w_{ij}>0$ if there is an edge from $i$
to $j$, and $w_{ij}=0$ otherwise.
We impose $w_{ii}=0$, so that if the graph contains
loops, we do not count them.
Node $i$ has out-degree
$w_{i+} = \sum_{j=1}^n w_{ij}$
and in-degree
$w_{+i} = \sum_{j=1}^n w_{ji}$.
The volume of the graph is $w_{++}=\sum_{i=1}^n\sum_{j=1}^n
w_{ij}$.\vspace{1,5pt}

There is a natural random walk associated with $w_{ij}$ in which
the probability of transition from $i$ to $j$ is $P_{ij} = w_{ij}/w_{i+}$.
Very often this walk is irreducible and aperiodic.
If not, it may be reasonable to modify $W$, by, for example,
adding a small probability of a transition uniformly distributed
on all nodes. For example, such a modification is incorporated
into the PageRank algorithm of \citet{Page} to yield an irreducible
and aperiodic walk on web pages.

An irreducible and aperiodic
walk has a unique stationary distribution, that we call $\pi$,
which places probability
$\pi_i$ on vertex $i$.
\citet{Zhou2005b} define the similarity between $i$ and $j$
to be $s_{ij} = \pi_iP_{ij}+\pi_jP_{ji}$.
Then they construct a variation functional for vectors $Z\in\mathbb{R}^n$
defined on the nodes of~$G$:
%
\begin{eqnarray}\label{eq:variation}
\Omega(Z) = \frac12\sum_{i,j} s_{ij}\biggl(
\frac{Z_i}{\sqrt{\pi_i}}-\frac{Z_j}{\sqrt{\pi_j}}
\biggr)^2.
\end{eqnarray}

This variation penalizes vectors $Z$ that differ
too much over similar nodes. It also contains a scaling
by $\sqrt{\pi_i}$. One intuitive reason for such a scaling
is that a small number of nodes with a large $\pi_i$ could reasonably
have more extreme values of $Z_i$, while the usually much greater
number of nodes with small $\pi_i$ should not ordinarily be allowed
to have very large $Z_i$, and hence should be regularized
more strongly.
Mathematically, the divisors $\sqrt{\pi_i}$ originate in
spectral clustering and graph partitioning algorithms.

The prediction $Z$ should have a small value of $\Omega(Z)$.
But it should also remain close to the observed values.
To this end, they make a vector $Y^*$ where $Y^*_i=y_i$
when $y_i$ is observed and $Y^*_i=\mu_i$ when $y_i$ is not
observed, where $\mu_i$ is a reasonable guess for $Y_i$.
Then the predictions are given by
%
\begin{eqnarray}\label{eq:rwsmoo}
\widehat Y = \operatorname{arg\,min}\limits_{Z\in\mathbb{R}^n} \Omega(Z) +
\lambda\Vert Z-Y^*\Vert^2,
\end{eqnarray}
where $\lambda>0$ is a parameter governing the trade off between
fit and smoothness.

The minimizer $\widehat Y$ in~\eqref{eq:rwsmoo} is a linear function
of $Y^*$.
In very many of the applications $y_i\in\{-1,1\}$ is
binary and $Y^*_i=0$ is used to represents uncertainty about their value.
Although linear models may seem less natural than logistic
regressions,
they are often used for discrete responses because they are a computationally
attractive relaxation of the problem of minimizing a quantity over
$Z\in\{-1,1\}^n$.

The outline of this paper is as follows.
Because the random walk smoother leads to a linear method,
we might expect it to have a representation
as a minimum mean squared error linear prediction
under a Gaussian process model for $Z$. That is, it
might be a form of kriging.
Section~\ref{sec:onkriging} presents notation
for kriging methods.
Section~\ref{sec:itskriging} exhibits a sequence
of kriging predictors that converge to
the random walk semi-supervised learning prediction~\eqref{eq:rwsmoo}.

The kriging model which yields random walk smoothing
has a covariance assumption
driven by the geometry of the graph, in which
the $\sqrt{\pi_i}$ values play a very strong role.
Section~\ref{sec:othersmoo} explores some other graph based
semi-supervised learning methods which have different covariance
assumptions in their corresponding kriging models.
In Section~\ref{sec:ourkriging} we derive another kriging
method incorporating the empirical variogram of the observed $Y_i$ values
into an estimate of the covariance.
That method uses a full rank covariance, which
is therefore computationally expensive for large $n$.
We also present a lower rank version more suitable to large
scale problems.

Section~\ref{sec:examples} presents two numerical
examples.
In Section~\ref{sec:unidata}
$Y_i$ is a numerical measure of the research quality
of $107$ universities in the UK and $w_{ij}$ measures
the number of links from university $i$ to $j$.
In holdout comparisons our kriging method is more
accurate than the random walk smoother, which
ends up being quite similar to a linear regression on $\sqrt{\pi_i}$
values without an intercept.
Section~\ref{sec:webkb} presents a binary example,
the WebKB data set, where the response is $1$ for
student web pages and $-1$ otherwise.
Incorporating empirical correlations
into the semi-supervised learning methods brings a
large improvement in the area under the ROC curve.

Section~\ref{sec:variation} describes some simple
adaptations of the approach presented here. Section~\ref{sec:otherlit}
discusses some related literature in fields other than machine learning.
Section~\ref{sec:conc} has our conclusions.

Our main contribution is two-fold. First, to the best of our knowledge,
we are the first to recognize the kriging framework underlying several
recently developed semi-supervised methods for data on graphs. Second,
we propose an empirical stationary correlation kriging algorithm which
adapts the traditional variogram techniques in Euclidean space to graphs.

\section{Kriging on a graph}\label{sec:onkriging}

Kriging is named for the mining engineer
Krige, whose paper~\citet{Krige} introduced
the method.
For background on kriging see
\citet{Stein} or \citet{cressie}.
Here we present the method
and introduce the notation we need later.

A plain model for kriging on a graph works as follows.
The data $Y\in\mathbb{R}^n$ are written
%
\begin{eqnarray}\label{eq:krigmodel}
Y = X\beta+ S + \varepsilon.
\end{eqnarray}
Here $X\in\mathbb{R}^{n\times k}$ is a matrix of
predictors and $\beta\in\mathbb{R}^k$
is a vector of coefficients.
The structured part of the signal is
$S\sim\mathcal{N}(0,\Sigma)$ and it is the correlations
within $\Sigma$ that capture how neighbors in
the graph are likely to be similar.
Finally, $\varepsilon\sim\mathcal{N}(0,\Gamma)$
is measurement noise independent of $S$.
The noise covariance $\Gamma$ is diagonal.

The design matrix $X$ has one row for each
node of the graph.
It can include a column of ones for an intercept,
columns for other predictors constructed from the
graph adjacency matrix $W$ and
columns for other covariates measured at the nodes.
To emphasize the role of the graph
structure, we only use covariates derived from~$W$.
At first we take
$X\in\mathbb{R}^n$, so $k=1$, and then make $\beta$
random from $\mathcal{N}(\mu,\delta^{-1})$, independent of
both $\varepsilon$ and $S$.
We write the result as
%
\begin{eqnarray}\label{eq:krigmodel2}
Y = Z + \varepsilon,
\end{eqnarray}
where $Z \sim\mathcal{N}(\mu X,\Psi)$, for $\Psi= \delta^{-1}XX' +
\Sigma$, independently of $\varepsilon\sim\mathcal{N}(0,\Gamma)$.

In this formulation the values $Y$ that we have observed
are noisy measurements of some underlying quantity $Z$
that we wish we had observed. We seek to recover $Z$
from measurements $Y$.

Some of the $Y_i$ are observed and some are
not. None of the $Z_i$ are observed.
We let $Y^{(0)}$ denote the random variables
that are observed, and $y^{(0)}$ be the values we saw for them.
The unobserved part of $Y$ is denoted by $Y^{(1)}$.
The kriging predictor is
%
\begin{eqnarray}\label{eq:krigExp}
\widehat Z = \mathbb{E}\bigl( Z\mid Y^{(0)}\bigr).
\end{eqnarray}

Without loss of generality, suppose that the vectors are ordered with
observed random variables before unobserved ones. We partition $\Psi$
as follows:
\[
\Psi=\left(
\matrix{ \Psi_{00} & \Psi_{01}\vspace*{2pt}\cr
 \Psi_{10} & \Psi_{11}
} \right)
 =
\left( \matrix{ \Psi_{\bullet0} & \Psi_{\bullet1} } \right) ,
\]
so that, for example, $\Psi_{00}=\operatorname{cov}( Z^{(0)},Z^{(0)})$
and $\Psi_{\bullet0}=\operatorname{cov}( Z,Z^{(0)})$. The
matrices~$\Sigma$ and $\Gamma$ are partitioned the same way.

The joint distribution of $Z$ and $Y^{(0)}$ is
\[
\pmatrix{ Z\vspace*{2pt}\cr Y^{(0)} }
\sim\mathcal{N}\left(
\pmatrix{ \mu X\vspace*{2pt}\cr \mu X^{(0)} }
,
\pmatrix{ \Psi& \Psi_{\bullet0} \vspace*{2pt}\cr \Psi_{0\bullet}& \Psi_{00} +
\Gamma_{00} }
\right),
\]
where $X^{(0)}$ contains the components of $X$ corresponding to
$Y^{(0)}$. Now we can write the kriging predictor \eqref{eq:krigExp}
explicitly as
%
\begin{eqnarray}\label{eq:krighat}
\widehat Z &=&
\Psi_{\bullet0}(\Psi_{00}+\Gamma_{00})^{-1}\bigl(y^{(0)}-\mu
X^{(0)}\bigr)+\mu X\nonumber
\\
&=& \bigl(XX^{(0)\prime}/\delta+ \Sigma_{\bullet0}\bigr)
\bigl(X^{(0)}X^{(0)\prime}/\delta+\Sigma_{00}+\Gamma
_{00}\bigr)^{-1}
\\
&&{}\times \bigl(y^{(0)}-\mu X^{(0)}\bigr)+\mu X.\nonumber
\end{eqnarray}
In the special case where
the whole vector $Y=y$ is observed, the kriging predictor is
%
\begin{eqnarray}\label{eq:krighatall}
\widehat Z = (XX'/\delta+ \Sigma) (XX'/\delta+\Sigma+\Gamma)^{-1}(y-\mu
X)+\mu X.
\end{eqnarray}

We have presented the kriging method under a Gaussian framework, where
estimators \eqref{eq:krighat} and \eqref{eq:krighatall} are the
conditional expectations and hence are the best predictors in terms of
minimizing mean squared error (MSE). However, even without the Gaussian
assumption, estimator \eqref{eq:krighat} and hence also
\eqref{eq:krighatall} is the best linear unbiased predictor (BLUP) of
$Z$, because it minimizes the MSE among all linear unbiased predictors
[see \citet{Stein}, Chapter 1]. For background on the BLUP, see
\citet{robi1991}.

Letting $\delta\to0$ leads to a model with an improper
prior in which $Y$ has infinite prior variance in the direction
given by $X$.
One natural choice for $X$ is the vector ${\mathbf{1}}_n$ of
all $1$s. Then the mean response over the whole graph is not penalized,
just fluctuations within the graph. We will see other natural
choices for $X$.

When these matrices are unknown, one can apply kriging by estimating
$\Sigma$, $\Gamma$~and $\mu$, and then predicting by
\eqref{eq:krighat}. The kriging approach also gives expressions for the
variance of the prediction errors:
%
\begin{eqnarray}\label{eq:krigvarest}
\operatorname{var}\bigl(Z\mid Y^{(0)}=y^{(0)}\bigr)= \Psi-
\Psi_{\bullet0}(\Psi_{00} + \Gamma_{00})^{-1}\Psi_{0\bullet}.
\end{eqnarray}

The predictions do not necessarily interpolate the known values.
That is, $\widehat Z^{(0)}$ need not equal $Y^{(0)}$.
Instead some smoothing takes place.
The predictions can
be forced closer to the data by making $\Gamma_{00}$ smaller.
One reason not to interpolate is that
when the graph correlations are strong,
it may be possible to detect erroneous labels
as cases where
$|\widehat Z^{(0)}_i -Y^{(0)}_i|$ is large.

\section{Random walk smoothing as kriging}\label{sec:itskriging}

Here we show that random walk regularization
can be cast in terms of a sequence of kriging
estimators.

The random walk regularizer predicts the responses $Y$ by
%
\begin{eqnarray}\label{eq:rwyhat}
\widehat Y = \operatorname{arg\,min}\limits_Z \frac12\sum_{i,j}s_{ij}
\biggl( \frac{Z_i}{\sqrt{\pi_i}}-\frac{Z_j}{\sqrt{\pi_j}}\biggr)^2
+\lambda\Vert Z-Y^*\Vert^2,
\end{eqnarray}
where $Y^*_i=Y_i$ for observed values and $Y^*_i=\mu_i$ for the
unobserved values. \citet{Zhou2005b} take $\mu_i=0$ for unobserved
$y_i\in\{-1,1\}$.

Introduce the matrix $\Pi= \operatorname{diag}(\pi_1,\dots,\pi_n)$,
let $s_{i+} = \sum_{j=1}^ns_{ij}$ and let
$\widetilde\Delta$ be the matrix with elements
\[
\widetilde\Delta_{ij} =
\cases{ s_{i+} -s_{ii}, &\quad $i=j$,\vspace*{2pt}\cr
 - s_{ij}, & \quad $i\ne j$. }
\]
The matrix $\widetilde\Delta$ is the graph Laplacian
of $\widetilde G$, which is our original graph $G$
after we replace the weights $w_{ij}$ by
the similarities $s_{ij}$. When $G$ is undirected,
the graph Laplacian $\Delta$ of $G$ is
\[
\Delta_{ij} =
\cases{ w_{i+} -w_{ii}, &\quad  $i=j$, \vspace*{2pt}\cr - w_{ij}, & \quad$i\ne j$. }
\]

It is clear that $\Delta$ and $\widetilde\Delta$ are symmetric matrices
with an eigenvalue of $0$ corresponding to eigenvector
${\mathbf{1}}_n$. Assuming that a graph such as $G$ or $\widetilde G$
is connected, its Laplacian is positive semi-definite and has rank
$n-1$ [\citet{Luxburg}]. For later use we write
%
\begin{eqnarray}\label{eq:deltaeig}
\widetilde\Delta= U' \operatorname{diag}(d_1,d_2,\dots,d_{n-1},0)
U=\sum
_{i=1}^nd_iu_iu_i^\prime,
\end{eqnarray}
where $U'U=I_n$, with $d_i>0$ for $i<n$ and $d_n=0$.

In matrix terms, the right-hand side of equation \eqref{eq:rwyhat} is
\begin{eqnarray*}
&& Z'\Pi^{-1/2}\widetilde\Delta\Pi^{-1/2}Z + \lambda (Z-Y^*)'(Z-Y^*)
\\
&&\qquad = Z'(\Pi^{-1/2}\widetilde\Delta\Pi^{-1/2} + \lambda I)Z
-2\lambda Z'Y^* +\lambda{Y^*}'Y^*.
\end{eqnarray*}
For $\lambda>0$ this is a positive definite quadratic
form in $Z$ and we find that
\begin{eqnarray}\label{eq:rwyhat2}
\widehat Y = \lambda(\Pi^{-1/2}\widetilde\Delta\Pi^{-1/2} + \lambda I
)^{-1}Y^*
=(I +
\lambda^{-1}\Pi^{-1/2}\widetilde\Delta\Pi^{-1/2})^{-1}Y^*.
\end{eqnarray}

Now we are ready to present the existing random walk algorithm as a
special form of kriging. To get the random walk predictor
\eqref{eq:rwyhat2}, we do the following:
\begin{enumerate}[(1)]
\item[(1)] make strategic choices for $\Gamma$, $\Sigma$ and $X$,
\item[(2)] treat the missing parts of $Y$ as observed,
\item[(3)] use the full data kriging estimator \eqref{eq:krighatall},
and then
\item[(4)] take the limit as $\delta\to0$ from above.
\end{enumerate}

In detail, the recipe is as follows:
\begin{theorem}\label{thm:itskriging}
Let $Y=Z+\varepsilon\in\mathbb{R}^n$.
Suppose that $Z=X\beta+S,$
where $X\in\mathbb{R}^n$, 
$\beta\sim\mathcal{N}(\mu,1/\delta)$ and $S\sim\mathcal{N}(0,\Sigma)$.
Let $\varepsilon\sim N(0,\Gamma)$ and assume that $S$, $\beta$ and
$\varepsilon$ are mutually independent. Suppose that $Y^{(0)}$
comprising the first $r>1$ elements of $Y$ is observed. Let
$Y^*\in\mathbb{R}^n$ with $Y^*_i = Y^{(0)}_i$ for $i=1,\dots,r$ and
$Y^*_i=\mu X_i$ for $i=r+1,\dots,n$. Let $\widehat Z^*_\delta$ be the
kriging estimator \eqref{eq:krighatall} applied with $y=Y^*$. Assume
that the Laplacian matrix $\widetilde\Delta$ derived for the similarity
weighted graph $\widetilde G$ on which $Y$ is defined satisfies
\eqref{eq:deltaeig}. We now choose
\begin{eqnarray*}
\Gamma &=& \lambda^{-1} I,
\\
\Sigma &=& \Pi^{1/2}\widetilde\Delta^+\Pi^{1/2},\quad\mbox{and}
\\
X&=& \bigl(\sqrt{\pi_1},\ldots,\sqrt{\pi_n}\bigr) ',
\end{eqnarray*}
where $\widetilde\Delta^+$ is the Moore--Penrose inverse of
$\widetilde\Delta$
and $\Pi= \operatorname{diag}(\pi_1,\dots,\pi_n)$.
Then
\[
\lim_{\delta \to  0^+} \widehat Z^*_\delta=
(I + \lambda^{-1}\Pi^{-1/2}\widetilde\Delta\Pi^{-1/2})^{-1}Y^*,
\]
which is the random walk predictor given by \eqref{eq:rwyhat2}.
\end{theorem}

\begin{pf}
First we notice that $X= \Pi^{1/2}{\mathbf{1}}_n$.
The kriging estimator is $\widehat Z^*_\delta= M_\delta(Y^*-\mu
X)+\mu
X$, where
%
\begin{eqnarray}\label{eq:mdelta}
M_\delta&=&\biggl(\frac{XX'}\delta+ \Sigma\biggr)
\biggl(\frac{XX'}\delta+ \Sigma+\Gamma\biggr)^{-1}\nonumber
\\
&=& \biggl(\frac{XX'}\delta+ \Pi^{1/2}\widetilde\Delta^+\Pi
^{1/2}\biggr) \biggl(\frac{XX'}\delta+
\Pi^{1/2}\widetilde\Delta^+\Pi^{1/2} +\lambda^{-1}I\biggr)^{-1}
\\
&=& \Pi^{1/2}\biggl(\frac{{\mathbf{1}}_n{\mathbf{1}}_n'}\delta+
\widetilde\Delta^+ \biggr)\Pi^{1/2}
\biggl(\Pi^{1/2}\biggl(\frac{{\mathbf{1}}_n{\mathbf{1}}_n'}\delta+
\widetilde\Delta ^+\biggr)\Pi^{1/2}
+\lambda^{-1}I\biggr)^{-1}.\nonumber
\end{eqnarray}

The three matrices on the left of \eqref{eq:mdelta} are invertible.
Moving their inverses inside the matrix inverse there, we get
\begin{eqnarray*}
M_\delta &=& \biggl(I
+\lambda^{-1}\Pi^{-1/2}\biggl(\frac{{\mathbf{1}}_n{\mathbf
{1}}_n'}\delta+ \widetilde \Delta^+\biggr)^{-1}\Pi^{-1/2}\biggr)^{-1}.
\end{eqnarray*}

Using \eqref{eq:deltaeig}, we write
\[
\frac{{\mathbf{1}}_n{\mathbf{1}}_n'}\delta+ \widetilde\Delta^+ =
U'\operatorname{diag}\biggl(\frac1{d_1},\frac1{d_2},\dots,\frac
1{d_{n-1}},\frac {n}\delta\biggr)U,
\]
noting that the last column of $U$ is $\pm{\mathbf{1}}_n/\sqrt{n}$,
the constant eigenvector. Now
\begin{eqnarray*}
M_\delta = \biggl(I +\lambda^{-1}\Pi^{-1/2}
U'\operatorname{diag}\biggl( d_1,d_2,\dots,d_{n-1},\frac\delta
{n}\biggr) U \Pi^{-1/2}\biggr)^{-1}.
\end{eqnarray*}
Letting $\delta\to0^+$,
\[
M_\delta\to M_0 = (I + \lambda^{-1}\Pi^{-1/2}\widetilde\Delta\Pi
^{-1/2})^{-1}.
\]
This limit exists because the matrix being inverted is positive definite.

The terms related to the mean $\mu X$ vanish because
\begin{eqnarray*}
(M_0X- X) &=& (I + \lambda^{-1}\Pi^{-1/2}\widetilde\Delta\Pi
^{-1/2})^{-1}\Pi^{1/2}{\mathbf{1}}_n- \Pi^{1/2}{\mathbf{1}}_n
\\
&=& (I + \lambda^{-1}\Pi^{-1/2}\widetilde\Delta\Pi
^{-1/2})^{-1}(\lambda ^{-1}\Pi^{-1/2}\widetilde\Delta{\mathbf{1}}_n+
\Pi^{1/2}{\mathbf {1}}_n) - \Pi
^{1/2}{\mathbf{1}}_n
\\
&=& (I + \lambda^{-1}\Pi^{-1/2}\widetilde\Delta\Pi ^{-1/2})^{-1}
\\
&&{}\times(\lambda ^{-1}\Pi^{-1/2}\widetilde\Delta\Pi^{-1/2} +
I)\Pi^{1/2}{\mathbf {1}}_n- \Pi ^{1/2}{\mathbf{1}}_n
\\
&=& 0.
\end{eqnarray*}
%
The second equality follows because $\widetilde\Delta{\mathbf{1}}_n=0$.
Therefore, in view of \eqref{eq:rwyhat2}, $\widehat Z_\delta^*\to
\widehat Y$ as $\delta\to0^+$.
\end{pf}

One thing that stands out from the kriging analysis is
the vector $X= \sqrt{\pi}$ interpreted component-wise.
The equivalent prior on $Y$ in the direction parallel to~$X$
is improper. Thus, the method anticipates that $Y$ could
reasonably be a large multiple of $X$.
When $Y\doteq\beta X$ for some value $\beta\ne0$,
the similar nodes are the ones with comparable
values of $\sqrt{\pi_i}$.
These are not necessarily close together in the graph.\looseness=1

The next thing that stands out is that the correlation
strength between nodes is a fixed property of $W$, the graph adjacency matrix.
If some response variables have stronger local correlations,
others weaker, and still others negative local correlations,
that is not reflected in this choice of $\Sigma$.

\section{Other semi-supervised learning as kriging}\label{sec:othersmoo}
There are several other graph based semi-supervised learning methods
that can
be expressed in a similar regularization framework.
In this section we build a similar connection between some of these
other semi-supervised learning methods and kriging. We state several
counterparts to Theorem \ref{thm:itskriging} but omit their proofs
because the details would be repetitive.
Most of these examples are taken from the survey paper
\citet{Zhu2005}. Some of these methods were originally introduced
with general loss functions, but we only consider their squared error
loss versions. This is because $\widehat Y$ may not be linear in $Y^*$
under a general loss function and hence is no longer kriging.

In each case there is a quadratic variation $\Omega(Z)$ and a
quadratic error
norm on $Z-Y^*$, each of which should ideally be small
subject to a tradeoff between them.
We take $\Omega(Z) = Z'LZ$ for a smoothing matrix $L$
and measure the error between $Z$ and $Y^*$ by $(Z-Y^*)'\Lambda
(Z-Y^*)$. The smoothing matrix $L$ is positive semidefinite and
$\Lambda$ is a diagonal matrix with $\Lambda_{ii}\geq0$, while the
sum $L+\Lambda$ is invertible.
The algorithm then picks the minimizer of
%
\begin{eqnarray}\label{eq:generalSemiCriterion}
Q(Z)= Z'LZ + (Z-Y^*)^\prime\Lambda(Z-Y^*).
\end{eqnarray}
It is easy to show that
%
\begin{eqnarray}\label{eq:generalSemiSol}
\widehat Y \equiv\arg\min_Z Q(Z) = (L+\Lambda)^{-1}\Lambda Y^*.
\end{eqnarray}
%
For random walk smoothing in Section \ref{sec:itskriging}, $\Lambda$ is
$\lambda I$ and $L$ is $\Pi^{-1/2}\widetilde\Delta\Pi^{-1/2}$.

Random walk smoothing is defined for directed graphs. A few of the
methods discussed below
are only defined for undirected graphs. To apply one of them to a given
directed graph,
the standard technique is to work with $W+W^\prime$.

We build the connection to kriging for several semi-supervised learning
methods below. Then, to allow easy comparison of the methods, we
present a summary in Table \ref{tab:choices} in Section \ref{sec5.1}.

\subsection{\texorpdfstring{Example one: Belkin, Matveeva and Niyogi (\protect\citeyear{2004Belkin})}
{Example one: Belkin, Matveeva and Niyogi (2004)}}
\label{sec:Tikhonov}

Belkin, Matveeva and Niyogi (\protect\citeyear{2004Belkin}) consider undirected graphs and use the (symmetric)
edge weights $w_{ij}$ as similarities $s_{ij}$. Their Tikhonov
regularization algorithm uses a criterion proportional to
\[
Q(Z) = Z'\Delta Z + \lambda_0\bigl\Vert Z^{(0)}-Y^{(0)}\bigr\Vert^2,
\]
where $\Delta$ is the graph Laplacian of $G$ (not $\widetilde G$).
They also have
an option to use the side constraint $\frac1n\sum_{i=1}^n Z_i =\frac
1r\sum_{i=1}^r Y_i^{(0)}$.
That constraint forces the mean prediction to equal the mean
observation, and is necessary for the generalization bound they obtained.
We do not use this condition, because the squared error\vspace*{1pt} norm on
$Z^{(0)}-Y^{(0)}$ already forces $Z^{(0)}$ to be close to $Y^{(0)}$.

Their method fits the quadratic criterion
\eqref{eq:generalSemiCriterion} after making the substitutions $L =
\Delta$ and $\Lambda= \operatorname{diag}(\lambda_0 I_r,  0 I_{n-r})$.
The solution $\widehat Y$ is the kriging estimator
\eqref{eq:krighatall} with the following choices:
\begin{eqnarray*}
\Gamma &=& \operatorname{diag}(\lambda_0^{-1}I_r, \lambda
_1^{-1}I_{n-r}),
\\
\Sigma &=& \Delta^{+},\quad\mbox{and }
\\
X &=& {\mathbf{1}}_n,
\end{eqnarray*}
taking the limit as $\delta\to0$ and then $\lambda_1 \to0$.

There are two key differences between this method and random walk smoothing.
First, neither $\Sigma$ nor $X$ involve $\sqrt{\pi}$ here.
Second, this model uses a diffuse prior on the noise for the unobserved
responses, while random walk smoothing uses the same variance for both
observed and unobserved responses. Therefore, this method avoids
plugging in a guess for the unobserved $Y^{(1)}$, and is thus
more typical of statistical practice.

\citet{2004Belkin} also propose an interpolating algorithm that leaves
all the known values unchanged in the prediction. That is, $\widehat
Y_i^{(0)}=Y_i^{(0)}$ for $i=1,\dots,r$. The resulting prediction arises
in the limit as $\lambda_0\to\infty$ for the Tikhonov estimator and,
hence, the connection to kriging remains the same.

They consider the generalization that replaces $\Delta$ by $\Delta^p$ for
a positive integer power $p$. They also consider a generalization in
which there could be more than one measurement made on the response
variable at some of the nodes.
We do not consider cases more general
than $0$ or $1$ observed response values per node.

\subsection{\texorpdfstring{Example two:
Zhou et al. (\protect\citeyear{Zhou2004})}{Example two: Zhou et al. (2004)}}

Zhou et al. (\citeyear{Zhou2004}) present an undirected graph algorithm that is a
predecessor to the random walk smoothing of \citet{Zhou2005b}. For
an undirected graph $w_{ij}=w_{ji}$, and of course the in- and
out-degrees of each node coincide. Let $D$ be the diagonal matrix
containing the common degree values $D_{ii} =w_{i+}=w_{+i}$.

They minimize
%
\begin{eqnarray}\label{eq:zhoufirst}
\frac12\sum_{i,j}w_{ij}\biggl(\frac{Z_i}{\sqrt{D_{ii}}} - \frac
{Z_j}{\sqrt{D_{jj}}}\biggr)^2 +\lambda\Vert Z-Y^*\Vert^2,
\end{eqnarray}
which is the random walk smoothing criterion \eqref{eq:rwyhat} after
replacing the similarity $s_{ij}$ by the weight $w_{ij}$ and the
stationary probability $\pi_i$ by the degree $D_{ii}$. Recall that for
an irreducible aperiodic random walk on an undirected graph with
transitions $P_{ij} = w_{ij}/w_{i+}$, the stationary distribution has
$\pi_i = D_{ii}/w_{++}$. Also, the similarity values become
proportional to $w_{ij}$: $s_{ij}=(D_{ii}/w_{++})(w_{ij}/D_{ii}) +
(D_{jj}/w_{++})(w_{ji}/D_{jj}) = 2w_{ij}/w_{++}$. As a result, the
symmetrized version of \eqref{eq:rwyhat} is equivalent to
\eqref{eq:zhoufirst} after multiplying $\lambda$ by $1/2$.

The criterion \eqref{eq:zhoufirst} fits the standard form
\eqref{eq:generalSemiCriterion}
with $L=D^{-1/2}\Delta D^{-1/2}$ and $\Lambda=\lambda I$,
where 
$\Delta$ is the graph Laplacian of $G$.

Their estimate reduces to the kriging estimator \eqref{eq:krighatall}
with the following choices:
\begin{eqnarray*}
\Gamma &=& \lambda^{-1} I,
\\
\Sigma &=& D^{1/2}\Delta^{+}D^{1/2},\quad\mbox{and}
\\
X&=& \bigl(\sqrt{D_{11}},\dots, \sqrt{D_{nn}}\bigr)^\prime,
\end{eqnarray*}
in the limit as $\delta\to0$.


\subsection{\texorpdfstring{Example three: Zhou, Sch\"{o}lkopf and
Hofmann (\protect\citeyear{Zhou2005a})}{Example three: Zhou, Sch\"{o}lkopf and Hofmann (2005)}}

Zhou,\break Sch\"{o}lkopf and Hofmann (\citeyear{Zhou2005a}) propose another random walk based strategy on
directed graphs that is motivated by the hub and authority web model
introduced by \citet{Kleinberg}. For \citet{Zhou2005a}, any
node with an outlink is a hub and any node with an inlink is an
authority. A node can be both a hub and an authority. They use two
random walks. Their hub walk transitions between hubs that link to a
common authority and their authority walk transitions between
authorities linked by a common hub.

The hubs define a walk on the authorities as follows.
From authority $i$ we pick a linking
hub $h$ with probability $w_{hi}/w_{+i}$ and from there pick an
authority $j$ with probability $w_{hj}/w_{h+}$.
The resulting transition probability from $i$ to $j$ is
\begin{eqnarray*}
P^{(A)}_{ij} = \sum_{h}\frac{w_{hi}}{w_{+i}}\cdot\frac{w_{hj}}{w_{h+}},
\end{eqnarray*}
where the sum is over hubs $h$.
Analogous hub transition probabilities are
\begin{eqnarray*}
P^{(H)}_{ij} = \sum_{a}\frac{w_{ia}}{w_{i+}}\cdot\frac{w_{ja}}{w_{+a}},
\end{eqnarray*}
summing over authorities $a$.

The stationary distributions of these two walks have closed forms
\begin{eqnarray*}
\pi_i^{(H)} =w_{i+}/w_{++} \quad\mbox{and}\quad \pi_i^{(A)}
=w_{+i}/w_{++}.
\end{eqnarray*}
These formulas give appropriate zeros for nodes $i$ that are
not hubs or authorities respectively.

We use stationary distributions and Laplacians of these two walks.
Let\vspace*{1pt} $\Pi_H=\operatorname{diag}(\pi_1^{(H)},\dots,\pi_n^{(H)})$
and $\Pi_A=\operatorname{diag}(\pi_1^{(A)},\dots,\pi_n^{(A)})$.
Then let $\widetilde\Delta_{H}$ be the\vspace*{1pt} Laplacian of the graph
$\widetilde G_H$,
which is our original graph $G$ after
replacing the weights $w_{ij}$ by the similarity $s_{ij}^{(H)}=\pi
_i^{(H)}P^{(H)}_{ij}+\pi_j^{(H)}P_{ji}^{(H)}$.
Similarly, let $\widetilde\Delta_A$ be the Laplacian of $\widetilde
G_A$ which has weights
$s_{ij}^{(A)}=\pi_i^{(A)}P^{(A)}_{ij}+\pi_j^{(A)}P^{(A)}_{ji}$.\vspace*{1pt}

The hub and authority regularization of \citet{Zhou2005a} uses the
quadratic criterion \eqref{eq:generalSemiCriterion} with $\Lambda=
\lambda I$ and smoothing matrix
%
\begin{eqnarray}\label{eq:hubauthL}
L = \gamma\Pi_H^{-1/2}\widetilde\Delta_H\Pi_H^{-1/2}
+ (1-\gamma)\Pi_A^{-1/2}\widetilde\Delta_A\Pi_A^{-1/2}
\end{eqnarray}
for some $\gamma\in[0,1]$. The choice of $\gamma$ allows the
user to weigh the relative importance of inlinks and outlinks.

Their hub and authority walk smoother matches the kriging estimator
\eqref{eq:krighatall} with the following choices:
\begin{eqnarray*}
\Gamma &=& \lambda^{-1} I,
\\
\Sigma &=& \bigl(\gamma\Pi_H^{-1/2}\widetilde\Delta_H\Pi_H^{-1/2} +
(1-\gamma)\Pi_A^{-1/2}\widetilde\Delta_A\Pi_A^{-1/2} \bigr)^{-1},\quad
\mbox{and}
\\
X &=& {\mathbf{0}}_n.
\end{eqnarray*}

Ordinarily, $L$ is positive definite for $0<\gamma<1$. The two terms in
\eqref{eq:hubauthL} each have one eigenvector with eigenvalue $0$, but
those two eigenvectors are, in general, linearly independent. We can
construct exceptions. For example, if $G$ is the complete graph, then
the hub and authority walks coincide and $L$ reduces to the random walk
case which has one zero eigenvalue. More generally, if every node has
$w_{i+}=w_{+i}$, the same thing happens. Outside of such pathological
examples, $L$ is positive definite.

\subsection{\texorpdfstring{Example four: Belkin, Niyogi and
Sindhwani (\protect\citeyear{Belkin2006})}{Example four: Belkin, Niyogi and Sindhwani (2006)}}
The manifold regularization framework introduced by
\citet{Belkin2006} considers undirected graphs with similarity
$s_{ij}=w_{ij}$. They predict the responses $Y$ by
\[
\widehat Y = \operatorname{arg\,min}\limits_Z \Vert
Z\Vert_\mathcal{K}^2 + \gamma Z^\prime\Delta Z + \lambda_0\bigl\Vert
Z^{(0)}-Y^{(0)}\bigr\Vert^2,
\]
where $\mathcal{K}$ is a Mercer kernel [see \citet{svmbook},
Chapter 3], $\Delta$ is the graph Laplacian and $\gamma>0$. The term
$\Vert Z\Vert_\mathcal{K}^2$ controls the smoothness of the predictions
in the \textit{ambient} space, while $Z^\prime\Delta Z $ controls the
smoothness with respect to the graph. We consider the special case
where $\mathcal{K}$ is a linear kernel. Then $\Vert Z\Vert_\mathcal{K}
^2=Z^\prime KZ$ for a positive semidefinite matrix
$K\in\mathbb{R}^{n\times n}$. Now manifold regularization uses the
criterion \eqref{eq:generalSemiCriterion} with $L=K+\gamma\Delta$ and
$\Lambda= \operatorname{diag}(\lambda_0 I_r,  0 I_{n-r})$.

We have two cases to consider.
The matrix $\gamma\Delta$ has $n-1$ positive eigenvalues
and an eigenvalue of $0$ for the eigenvector ${\mathbf{1}}_n$.
If $K{\mathbf{1}}_n= 0$, then $L=K+\gamma\Delta$ is singular, but
otherwise $L$ is positive definite.

When $L$ is positive definite, the implied prior is not improper in any
direction so we take $X={\mathbf{0}}_n$. In this case, the manifold
regularization predictions are from the kriging estimator
\eqref{eq:krighatall} with the following choices:
\begin{eqnarray*}
\Gamma &=&
\pmatrix{ \lambda_0^{-1}I_r & 0\vspace*{2pt}\cr 0 &\lambda_1^{-1}I_{n-r} }
,\\
\Sigma &=& (K + \gamma\Delta)^{-1},\quad\mbox{and}
\\
X &=& {\mathbf{0}}_n,
\end{eqnarray*}
in the limit $\lambda_1 \to0$.

Now suppose that $K +\gamma\Delta$ fails to be invertible
because $K$ has eigenvector ${\mathbf{1}}_n$ with eigenvalue $0$. In
this case, we replace $(K+\gamma\Delta)^{-1}$ by the corresponding
Moore--Penrose inverse and use
$X={\mathbf{1}}_n$, taking the limit $\delta\to0$.

Our condition that the Mercer kernel be linear is necessary.
For a general Mercer Kernel $\mathcal{K}$, the prediction $\widehat Y$
need not be
linear in $Y^*$,
and so for such kernels, manifold regularization does not reduce to kriging.

\subsection{Other examples: smoothing matrix derived from $\Delta$}

A few papers [e.g., \citet{Kondor}, \citet{Smola},
\citet{Zhu2003}] construct the smoothing matrix $L$ based on a
spectral transformation of the graph Laplacian $\Delta$. They take
\[
L = \sum_{i=1}^n f(d_i)u_iu_i^\prime,
\]
where $d_i$ and $u_i$ are eigenvalues and eigenvectors of $\Delta$ as
in \eqref{eq:deltaeig}, and $f(\cdot)$ is a nonnegative increasing
function, such as $f(x)=e^{\alpha^2x/2}$.

When $f(d_n)>0$, the connection to kriging can be written as
\begin{eqnarray*}
\Gamma &=&
\pmatrix{ \lambda_0^{-1}I_r & 0\vspace*{2pt}\cr 0 &\lambda_1^{-1}I_{n-r} }
,\\
\Sigma &=& \sum_{i=1}^n f(d_i)^{-1}u_iu_i^\prime,\quad\mbox{and}
\\
X &=& {\mathbf{0}}_n.
\end{eqnarray*}
For $f(d_n)=0$, $\Gamma$ remains the same but now
\begin{eqnarray*}
\Sigma &=& \sum_{i=1}^{n-1} f(d_i)^{-1}u_iu_i^\prime \quad\mbox{and}
\\
X &=& {\mathbf{1}}_n,
\end{eqnarray*}
with $\delta\to0$.

\section{Empirical stationary correlations}\label{sec:ourkriging}

In the last two sections we have established connections between
kriging and several semi-supervised learning models for prediction on
graphs. Such relationships are themselves interesting, but what is more
striking to us is that the connections to kriging reveal a unanimous
assumption by all these models: the signal covariance is a given
function of the graph adjacency matrix $W$. This is clearly not an
effective way to capture the various correlation properties that
different response variables may present. For instance, even on the
same social network (i.e., the same $W$), friends may correlate
differently for age, than for gender, school attended or opinions about
music, movies or restaurants.

This troubling feature motivates us to propose a different model for
the signal covariance that can adapt to the nature of the response
variable. Similar to the prediction methods discussed thus far, we keep
the kriging framework as presented in Section \ref{sec:onkriging}, but
now we show how to adapt the covariance to the dependency pattern seen
among the non-missing $Y$ values.



\subsection{Stationary correlations}\label{sec5.1}

We start with the model for the underlying signal~$Z,$
%
\begin{eqnarray}\label{eq:statempcorr}
Z\sim\mathcal{N}( \mu X,   \sigma^2VRV),
\end{eqnarray}
where the covariance is decomposed into a correlation matrix $R\in
\mathbb{R}
^{n\times n}$, a diagonal matrix $V=\operatorname{diag}(v_1,\dots
,v_n)$ containing
known relative standard deviations $v_i>0$, and a scale parameter
$\sigma>0$.

Model \eqref{eq:statempcorr} includes both random walk regularization
(Section \ref{sec:itskriging}) and the Tikhonov regularization (Section
\ref{sec:Tikhonov}) as special cases. The connections are made in Table
\ref{tab:summaryTable}.

\begin{table}[b]
\tablewidth=193pt
 \caption{Parameters chosen for model
\protect\eqref{eq:statempcorr} to obtain the random walk smoothing and
the Tikhonov smoothing methods. Both models use the limit
$\delta\to0$}\label{tab:summaryTable}
\begin{tabular*}{193pt}{@{\extracolsep{4in minus 4in}}lccc@{}}
\hline
& $\bolds{X}$ & $\bolds{v}$ & $\bolds{\sigma^2R_{ij}}$
\\
\hline
Random walk: & $\sqrt{\pi}$ & $\sqrt{\pi}$&
$\widetilde\Delta^+_{ij}+\delta^{-1}$
\\[3pt]
Tikhonov: & ${\mathbf{1}}_n$ & ${\mathbf{1}}_n $ &
$\Delta^+_{ij}+\delta^{-1}$
\\
\hline
\end{tabular*}
\end{table}

\begin{table}[t]
\tabcolsep=0pt \caption{Summary of connections between some
semi-supervised learning methods and kriging} \label{tab:choices}
\begin{tabular*}{\tablewidth}{@{\extracolsep{4in minus 4in}}lcccc@{}}
\hline \textbf{Reference} & $\bolds{\Gamma}$ & $\bolds{\Sigma}$ &
$\bolds{X}$ & \textbf{Limits}
\\
\hline
 \citet{Zhou2005b} & \multicolumn{1}{c}{\multirow{2}{20pt}{\centering{{$\lambda^{-1} I$}}}} &
\multicolumn{1}{c}{\multirow{2}{52pt}{\centering{{$\Pi^{1/2}\widetilde\Delta^{+}\Pi^{1/2}$}}}} &
\multicolumn{1}{c}{\multirow{2}{28pt}{\centering{{$\Pi^{1/2}{\mathbf{1}}_n$}}}} &
\multicolumn{1}{c}{\multirow{2}{33pt}{\centering{{$\delta\to0$}}}}
\\
\quad (Random walk) & & & &\\
\citet{2004Belkin} &
\multicolumn{1}{l}{\multirow{2}{66pt}{\centering
{\fontsize{7.6}{7.6}\selectfont$\left(\matrix{\lambda_0^{-1}I_r
& 0\vspace*{2pt}\cr 0 & \lambda_1^{-1}I_{n-r}}\right)$}}} &
\multicolumn{1}{c}{\multirow{2}{14pt}{{\centering $\Delta^{+}$}}} &
\multicolumn{1}{c}{\multirow{2}{9pt}{{\centering ${\mathbf{1}}_n$}}} &
\multicolumn{1}{c}{\multirow{2}{33pt}{\centering $\delta\to0$
$\lambda_1 \to0$}}
 \\
\quad (Tikhonov) & & & &
\\
\citet{2004Belkin} &\multicolumn{1}{l}{\multirow{3}{66pt}{\centering
{\fontsize{7.6}{7.6}\selectfont$\left(\matrix{\lambda_0^{-1}I_r
& 0\vspace*{2pt}\cr 0 & \lambda_1^{-1}I_{n-r}}\right)$}}}  & \multicolumn{1}{c}{\multirow{3}{14pt}{\centering
$\Delta^{+}$}} & \multicolumn{1}{c}{\multirow{3}{9pt}{\centering
${\mathbf{1}}_n$}} & \multicolumn{1}{c}{\multirow{3}{33pt}{\centering
$\delta\to0$ $\lambda_1 \to0$ $\lambda_0 \to\infty$}}
\\
\quad (Interpolated) & & & &
\\
& & & &
\\
\citet{Zhou2004} & $\lambda^{-1} I$ & $D^{1/2}\Delta^{+}D^{1/2}$ &
$D^{1/2}{\mathbf{1}}_n$ & $\delta\to0$
\\
\citet{Zhou2005a} & \multicolumn{1}{c}{\multirow{2}{20pt}{\centering{{$\lambda^{-1} I$}}}} &
\multicolumn{1}{c}{\multirow{2}{79pt}{\centering{\fontsize{7.6}{7.6}\selectfont$
\matrix{((1-\gamma)\Pi_A^{-1/2}\widetilde\Delta _A\Pi _A^{-1/2} \vspace*{2pt} \cr +
\gamma\Pi_H^{-1/2}\widetilde\Delta_H\Pi _H^{-1/2})^{-1} } $}}} &
\multicolumn{1}{c}{\multirow{2}{9pt}{\centering{{${\mathbf{0}}_n$}}}} &
\multicolumn{1}{c}{\multirow{2}{33pt}{\centering{--}}}
\\
\quad (Hub \& authority) & & & &
\\
\citet{Belkin2006} &
\multicolumn{1}{l}{\multirow{2}{66pt}{\centering
{\fontsize{7.6}{7.6}\selectfont$\left(\matrix{\lambda_0^{-1}I_r
& 0\vspace*{2pt}\cr 0 & \lambda_1^{-1}I_{n-r}}\right)$}}}  &
\multicolumn{1}{c}{\multirow{2}{49pt}{\centering{$(K+\gamma\Delta)^{-1}$}}}
& \multicolumn{1}{c}{\multirow{2}{9pt}{\centering{${\mathbf{0}}_n$}}}
& \multicolumn{1}{c}{\multirow{2}{28pt}{\centering{{$\lambda_1 \to0$}}}}
\\
\quad (Manifold, $K{\mathbf{1}}_n\neq{\mathbf{0}}_n$) & & & &
\\
\citet{Belkin2006} &
\multicolumn{1}{l}{\multirow{2}{66pt}{\centering
{\fontsize{7.6}{7.6}\selectfont$\left(\matrix{\lambda_0^{-1}I_r
& 0\vspace*{2pt}\cr 0 & \lambda_1^{-1}I_{n-r}}\right)$}}} &
\multicolumn{1}{c}{\multirow{2}{49pt}{\centering{{$(K+\gamma\Delta)^{+}$}}}}
& \multicolumn{1}{c}{\multirow{2}{9pt}{\centering{{${\mathbf{1}}_n$}}}}
& \multicolumn{1}{c}{\multirow{2}{28pt}{\centering{$\delta\to0$
$\lambda_1\to 0$}}}
\\
\quad (Manifold, $K{\mathbf{1}}_n={\mathbf{0}}_n$) & & & &
\\
Spectral transform &
\multicolumn{1}{l}{\multirow{2}{66pt}{\centering
{\fontsize{7.6}{7.6}\selectfont$\left(\matrix{\lambda_0^{-1}I_r
& 0\vspace*{2pt}\cr 0 & \lambda_1^{-1}I_{n-r}}\right)$}}} &
\multicolumn{1}{c}{\multirow{2}{70pt}{\centering{{$\sum_{i=1}^{n}
f(d_i)^{-1}u_iu_i^\prime$}}}} &
\multicolumn{1}{c}{\multirow{2}{9pt}{\centering{{${\mathbf{0}}_n$}}}} &
\multicolumn{1}{c}{\multirow{2}{28pt}{\centering{--}}}
\\
\quad $f(d_n)>0$ & & & &
\\
Spectral transform &
\multicolumn{1}{l}{\multirow{2}{66pt}{\centering
{\fontsize{7.6}{7.6}\selectfont$\left(\matrix{\lambda_0^{-1}I_r
& 0\vspace*{2pt}\cr 0 & \lambda_1^{-1}I_{n-r}}\right)$}}} &
\multicolumn{1}{c}{\multirow{2}{72pt}{\centering{{$\sum_{i=1}^{n-1}
f(d_i)^{-1}u_iu_i^\prime$}}}} &
\multicolumn{1}{c}{\multirow{2}{9pt}{\centering{{${\mathbf{1}}_n$}}}} &
\multicolumn{1}{c}{\multirow{2}{28pt}{\centering{{$\delta\to0$}}}}
\\
\quad $f(d_n)=0$ & & & &
\\
\hline
\end{tabular*}
\end{table}

The key element in \eqref{eq:statempcorr} is the correlation matrix
$R$. All of the methods summarized in Table \ref{tab:choices} take $R$
to be a fixed matrix given by the graph adjacency matrix, via $\Delta$,
$\widetilde\Delta$, $\Pi$ and related quantities. Our model for
$R_{ij}$ is $\rho(s_{ij})$, where $\rho$ is a smooth function to be
estimated using the response values, and $s_{ij}$ is a measure of graph
similarity. These correlations are stationary in $s$, by which we mean
that two node pairs $ij$ and $i'j'$ from different parts of the graph
are thought to have the same correlation, if $s_{ij}=s_{i'j'}$. The
standard deviations $\sigma v_i$, by contrast, are proportional to
given numbers that need not be stationary with respect to any feature
of the nodes. The signal means need not be stationary either.
The estimation procedure, including measures to make $R$ positive
semidefinite, is described in detail in Section \ref{sec:estimation}
below.

Like these regularization methods, we assume in model \eqref
{eq:statempcorr} that $X$, $v$ and $s_{ij}$ are prespecified based on
domain knowledge. We take the noise variance to be $\lambda^{-1}I$,
like the random walk does, but unlike the Tikhonov method, which uses
effectively infinite variance for the unmeasured responses.

In the examples in Section \ref{sec:examples}, when we compare to the
random walk method, we will use $s_{ij}=\pi_iP_{ij}+\pi_jP_{ji}$.
Similarly, when we compare to the Tikhonov regularized method, we will
use $s_{ij} = w_{ij}$, or symmetrized to $w_{ij}+w_{ji}$ if the graph
is directed. In this way we will use the exact same similarity measures
as those methods do. There is one additional subtlety. We found it more
natural to make the correlation a smooth function of similarity. For
the other methods it is the inverse of the correlation that is smooth
in $s_{ij}$.

In matrix form, 
our prediction is 
%
\begin{eqnarray}\label{eq:krighat2}
\widehat Z = \Psi_{\bullet0}(\Psi_{00}+\lambda
^{-1}I)^{-1}\bigl(y^{(0)}-\mu X_0\bigr)+\mu X,
\end{eqnarray}
where $\Psi=\sigma^2VRV$, and we use the estimate $R$ described next.

\subsection{Covariance estimation through the variogram}\label{sec:estimation}


Here we adapt the vario\-gram-based approach from geostatistics [see,
for example, \citet{cressie}] to estimate the matrix $R$. With noise
variance $\lambda^{-1} I$, the variogram
of the model \eqref{eq:statempcorr}
is
\begin{eqnarray}\label{eq:variogram}
\Phi_{ij} &\equiv& \tfrac{1}{2}\mathbb{E}\bigl((Y_i-\mu X_i)-(Y_j-\mu
X_j) \bigr)^2\nonumber
\\[-8pt]\\[-8pt]
&=&\lambda^{-1}+\tfrac{1}{2}\sigma^2(v^2_i+v^2_j-2v_iv_j
R_{ij}).\nonumber
\end{eqnarray}
For $1\le i,j\le r$ both $Y_i=y_i$ and $Y_j=y_j$ are observed and so
we have the naive estimator
%
\begin{eqnarray}\label{eq:variogramEst}
\widehat\Phi_{ij} = \tfrac{1}{2}\bigl((y_i-\mu X_i)-(y_j-\mu
X_j)\bigr)^2.
\end{eqnarray}

The naive variogram is our starting point. We translate it into a naive
value $\widehat R_{ij}$ by solving equation \eqref{eq:variogram}.
This requires the prespecified values of $v_i$ and $v_j$.
We also need values for $\lambda$
and $\sigma$, which we consider fixed for now and will discuss
how to choose them later.

Once we have the naive correlation estimates $\widehat R_{ij}$, we
use a spline smoother to fit the smooth function $\widehat R_{ij}
\doteq
\hat\rho(s_{ij})$.
Smoothing serves two purposes.
It yields correlation as a function of similarity $s_{ij}$, and
it reduces sampling fluctuations.
Next we use $\hat\rho$ to estimate the entire correlation
matrix via $\widetilde R_{ij} = \hat\rho(s_{ij}),$ for $i\ne j$
with of course $\widetilde R_{ii}=1$.
To complete our estimation of the signal variance,
we take $\widehat\Psi= \sigma^2 V\widetilde RV$, and then if necessary
modify $\widehat\Psi$ to be positive semi-definite. Two versions of the
last step are considered. One is to use $\widehat\Psi_+$, the positive
semi-definite matrix that is closest to $\widehat\Psi$ in the Frobenius
norm. The other method is to use a low rank version of $\widehat\Psi
_+$ to
save computational cost.

The step-by-step procedure to estimate the signal covariance is listed
in Table~\ref{tab:covEst}.
The output is the estimated $\Psi$, which we use through equation
\eqref{eq:krighat2} to make predictions.

We choose $\sigma$ and $\lambda$ by cross-validation. This is the same
technique used by the semi-supervised methods discussed in Sections
\ref{sec:itskriging} and \ref{sec:othersmoo}. It is also similar to
treatment of the shrinkage parameter used in ridge
regression.%

In our cross-validation, some known labels are temporarily treated as
unknown and then predicted after fitting to the other labels. The
entire graph structure is retained, as that mimics the original
prediction problem. When estimating error rates we use training, test
and validation subsets.


\begin{table}[t]
 \caption{ The steps we use to estimate the covariance matrix
$\Psi=\sigma^2VRV$ in model \protect\eqref{eq:statempcorr} via an
empirical stationary correlation model} \label{tab:covEst}
\begin{tabular*}{\tablewidth}{@{\extracolsep{4in minus 4in}}p{1.0\linewidth}@{}}
\hline \textbf{Variance estimation with stationary correlations}
\\
\hline
Given $\lambda>0$ and $\sigma>0$:
\begin{enumerate}[1.]
\item For every pair of observed nodes $i,j=1,\dots,r$ and $i\neq j$,
estimate $R_{ij}$ by solving \eqref{eq:variogram} with $\Phi_{ij}$
estimated using \eqref{eq:variogramEst}:
%
\begin{equation}\label{eq:rawRij}
\widehat R_{ij}=\frac{\sigma^2(v^2_i+v^2_j)/2 + \lambda^{-1} -
\widehat\Phi
_{ij}}{\sigma^2v_iv_j}.
\end{equation}
\item Smooth the pairs $\{(\widehat R_{ij},s_{ij})\dvtx
i,j=1,\dots,r\}$ to obtain the estimated correlation function
$\hat{\rho}(\cdot)$.
\item Compute $\widetilde R_{ij}=\hat{\rho}(s_{ij})$ for $i\neq j$
and $\widetilde R_{ii}=1$.
\item Set $\widehat\Psi=\sigma^2V\widetilde RV$.
\item Use one of the following two methods to make $\widehat\Psi$
positive semi-definite. Let $\widehat\Psi=U^\prime HU$ be the
eigen-decomposition of $\widehat\Psi$. Then
\begin{enumerate}[(a)]
\item[(a)] use $\widehat\Psi_+=U^\prime H_+ U$, where $H_+=\max(H,0)$,
or,
%
\item[(b)] use 
$\widehat\Psi_+^{(k)}=U^\prime H_+^{(k)} U$, where $H_+^{(k)}$ consists
of the first $k$ diagonal elements of $H_+$ and the rest are set to be
zero.
\end{enumerate}
Choice (a)
gives the positive semi-definite matrix that is closest to $\widehat
\Psi$
in the Frobenius norm.
Choice (b) is used when
computational cost is a concern or
the true covariance $\Psi$ is believed to be
low-rank.
\end{enumerate}
\\[-10pt]
\hline
\end{tabular*}
\end{table}

\subsection{Practical issues}\label{sec:practical}
As we have discussed, we need to make choices for $X$, $v$ and $s_{ij}$
that go into our model \eqref{eq:statempcorr}. These prespecified
values should come from domain knowledge about the response variable of
interest. They may depend on the graph adjacency matrix $W$, but not on
the realization of the variable $Y$. The single predictor $X$
corresponds to the direction that $Y$ varies along and $v$ the amount
of univariate variations. The similarity $s_{ij}$ defines the closeness
of nodes $i$ and $j$. There are clearly many possible choices for these
parameter values, and we
do not yet have any guidance on how best to select them for a specific problem.


 The connections to the random walk and the Tikhonov smoothing
methods present two sets of example choices, as listed in Table
\ref{tab:summaryTable}. While $v$ and $X$ can be set separately, both
methods take $v=X$, and so signals $Z_i$ with a larger absolute mean
$|\mu X_i|$ also have a larger variance $\sigma^2X_i^2$. This seems
reasonable but of course some data will be better fit by other
relationships between mean and variance. One appealing feature of
choosing $v=X$ is that \eqref{eq:statempcorr} has a simple
interpretation that the scaled signals $Z_i/X_i$ are Gaussian with
constant mean, constant variance, and stationary correlation $R$. We
will compare the two sets of choices using real examples in Section
\ref{sec:examples} below.


It is worthwhile to point out that, for unweighted graphs, the Tikhonov
smoothing method leads to very few distinct values of $s_{ij}$. In this
case, we simply use the average of $\widehat R_{ij}$ for each distinct
$s_{ij}$ to
approximate $\hat\rho$ without smoothing. For choices that lead to many
distinct values of $s_{ij}$, cubic splines with ten knots are used to
get $\hat\rho$ out of convenience.
Better choices of smoothing method
could probably be made, but we expect the differences among
adaptive correlation methods to be smaller than those between
methods with fixed correlations and methods with simple
adaptive correlations.

Finally, all measurement errors have the same variance $\lambda^{-1}$
in our model.
It is unnecessary to assume a different noise variance for the
unobserved $Y$ because
their variance does not affect our predictor \eqref{eq:krighat2}.


\subsection{Relation to geostatistics}

We use a nonparametric estimate $\rho(\cdot)$ to avoid
forcing a parametric shape on the correlation function.
The parametric
curves used in the geostatistics literature for $\mathbb{R}^d$ with
small $d$ may not extend naturally to graphs, even if
they could be properly embedded in Euclidean space.

Both \citet{Hall} and \citet{Shapiro} have discussed ways to
fit a nonparametric variogram while ensuring a positive semi-definite
covariance. Their techniques apply when the predictor space is
$\mathbb{R}^d$. The usual definition of the similarity measure on a
graph is far from being a metric in~$\mathbb{R}^d$. Our approach
ensures that the estimate for $\Psi$ is positive semi-definite.

When there are $n$ observations, \citet{Hall} find convergence
rates for the smoother $\hat\rho$ that are comparable to that using
$n^2$ observations. The reason is that we get $O(n^2)$ pairs
$(Y_i,Y_j)$ in the empirical variogram. In our application there are
only $r(r-1)/2$ observed pairs to use.

In the spatial smoothing problems where kriging
originates, it is often necessary for
the covariance to remain semi-definite at any finite list of
points in $\mathbb{R}^d$, including some that are not yet observed.
Our setting does not require us to construct an
extension of the covariance function to $Y_i$
for nodes $i$ that are not in the graph.
Even in cross-validation, we know the positions in
the graph for the points at which no $Y$ values have been
observed, and so we can still compute $s_{ij}$
for all data pairs. This aspect of the semi-supervised setting
makes the problem much simpler than that faced in
geostatistics. It does, however, mean that when the
graph changes, the covariance model may have to be recomputed.

\section{Examples}\label{sec:examples}

In this section we compare our empirical covariance approach to the
random walk and the Tikhonov regularization methods. We use two
extremely different real data sets. The first one has a continuous
response on a dense, weighted graph, and the second one has a binary
response on a sparse, unweighted graph. Because both graphs are
directed, we construct an undirected graph for the Tikhonov approach
using $W+W^\prime$ as the adjacency matrix. Our empirical-based
method, together with its low rank variations,
brings substantial improvements for both methods on both data sets.

Recall that we need to prespecify the values of $X$, $v$ and $s_{ij}$
for our empirical covariance approach. Following the discussion in
Section \ref{sec:practical}, we consider two versions of our model. One
follows the choices of the random walk method and the other follows the
Tikhonov method, which we call ``empirical random walk'' and
``empirical Tikhonov'' respectively, compared to the ``original random
walk'' and the ``original Tikhonov.'' Therefore, the comparisons using
real data serve two purposes. First, the comparison between the
empirical and the original shows how performance changes when we
incorporate empirical stationary correlations. Second, the comparison
between the two original (or empirical) methods shows how the choices
of the prespecified parameters affect the predictions.

We also need to estimate the overall mean $\mu$.
For binary problems with $Y_i\in\{-1,1\}$, we take $\mu=0$, as is
done in the machine learning literature.
For continuous responses we use
%
\begin{eqnarray}\label{eq:muhat}
\hat{\mu}=\frac{1}{r}\sum_{i=1}^r\frac{y_i}{X_i}.
\end{eqnarray}
We also investigated estimating $\mu$ by generalized least squares
regression of $Y^{(0)}$ on $X^{(0)}$, taking account of estimated
correlations among the first $r$ response values. This made only a very
small difference even on the small problems we are about to report, and
so we see no reason to prefer it to the very simple estimate
\eqref{eq:muhat}. We do want to point out that estimating $\mu$ is
necessary for the empirical methods and the original random walk
method, but not necessary for the original Tikhonov method. This is
because even though $\mu$ is used in the construction of $Y^*$, it
disappears from $Y^*$ in the $\lambda_1\to0$ limit for the original
Tikhonov method.


\setcounter{footnote}{3}
\subsection{The UK university web link data set}\label{sec:unidata}
\subsubsection*{Data description}
The university data set contains the number of web links between UK
universities in 2002. Each university is associated with a research
score (RAE), which measures the quality of the university's
research.\footnote{The data are at
\url{http://cybermetrics.wlv.ac.uk/database/stats/data/}. We use the
link counts at the directory level.} After removing four universities
that have missing RAE scores, or that have no in-link or out-link,
there are 107 universities.

The response variable, RAE score, is continuous and ranges from $0.4$
to $6.5$ with a mean of $3.0$ and a variance of $3.5$. The number of
links from one university to another forms the (asymmetric) weighted
adjacency matrix $W$. The distribution of the weights $w_{ij}$ is
heavily right tailed and approximately follows a power law. About $15\%
$ of the weights are zero, and $50\%$ of them are less than $7$, while
the maximum is $2130$.

\subsubsection*{Illustration of the empirical covariance method}
We first use the entire data set to illustrate the empirical variance
estimation procedure as given in Table \ref{tab:covEst}. We illustrate
only the empirical Tikhonov method and hence use $v=X={\mathbf{1}}_n$
and $s_{ij}=w_{ij}+w_{ji}$. These similarity scores take many values,
and so we use correlation smoothing. The empirical random walk method
is similar.
In practice, $\sigma^2$ and $\lambda$ are chosen by cross-validation,
but we fix $\sigma^2=5$ and $\lambda^{-1}=0.01$ here to show one
iteration of the estimation procedure.

\begin{figure}[t]

\includegraphics{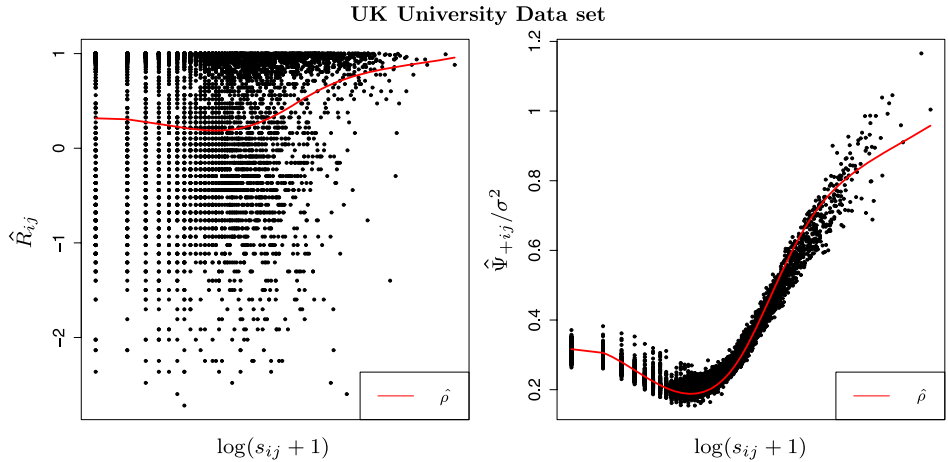}

\caption{Illustration of the empirical Tikhonov method with
the UK university data. Left: scatter plot of the naive $\widehat
R_{ij}$ values versus $\log(s_{ij}+1)$ with the cubic spline smoothing
curve (red). Right: final estimates $\widehat\Psi_{+ij}/\sigma^2$
versus $\log(s_{ij}+1)$ with the same smoothing curve (red).}
\label{uniRho}
\end{figure}

Figure \ref{uniRho} (left) plots the naive estimates $\widehat R_{ij}$,
as computed in \eqref{eq:rawRij}, against (log transformed) similarity
$s_{ij}$ values. The logarithm is used because the $s_{ij}$ are skewed.
The scatter plot is very noisy, but we can nonetheless extract a
nontrivial $\hat{\rho}(\cdot)$ with cubic spline smoothing (ten knots),
as shown by the red curve.
The same curve is also included on the right plot at a larger scale.

It is striking that $\hat{\rho}(\cdot)$ is not monotonically
increasing in $s_{ij}$.
The greatest correlations arise for very similar nodes, but the very
least similar node pairs
also have somewhat more correlation than do pairs with intermediate
similarity. Recall that a similarity of $0$ means that the pair of universities
are not linked. Universities without many links are overrepresented in such
pairs, and those universities tend to have similar (low) RAE scores.

The final step in Table \ref{tab:covEst} is to make the covariance
matrix $\widehat\Psi$ that directly results from $\hat{\rho}(\cdot)$
positive semi-definite. For the full rank version $\widehat\Psi_+$, we
plot points $\widehat\Psi_+/\sigma^2$ on the right side of Figure
\ref{uniRho}. These scatter around the red curve which shows
$\hat\rho$. We saw similar patterns (not shown here) with some low rank
estimates $\widehat\Psi_+^{(k)}$. During this final step, we saw in
Figure \ref{uniRho} (right) that a small number of highly similar node
pairs got the greatest change in model correlation. That pattern did
not always arise in other examples we looked at.

\subsubsection*{Performance comparisons}
Now we turn to performance comparisons.
For this, we hold out the RAE scores of some
universities and measure
each prediction method by
mean squared error (MSE) on the held out scores. The size of the
holdout set ranges from approximately $10\%$ to $90\%$ of the entire
data set, and 50 trials are done at each holdout level.

Our empirical methods have two tuning parameters $\lambda$ and $\sigma
$, while the original random walk and Tikhonov methods have only one.
Nevertheless, the comparison is fair because
it is based on holdout sets. For each set of held-out data
we used ten-fold cross-validation within the held-in data to pick
$\lambda$ and $\sigma$
for empirical stationary correlation kriging. For the original random
walk and Tikhonov
methods we use
the \textit{best} tuning parameter ($\lambda$), and so our comparisons
are to somewhat better versions of the random walk and Tikhonov method
than one could actually get in practice.

We define a baseline method that considers no signal covariance, and
simply regresses the responses $Y_i$ on $X_i$. With the random walk
choice of $X_i=\sqrt{\pi_i}$, the baseline prediction is $\hat\mu
\sqrt{\pi_i}$, while with the Tikhonov choice of $X_i=1$, it is
simply $\hat\mu$.

\begin{figure}[t]

\includegraphics{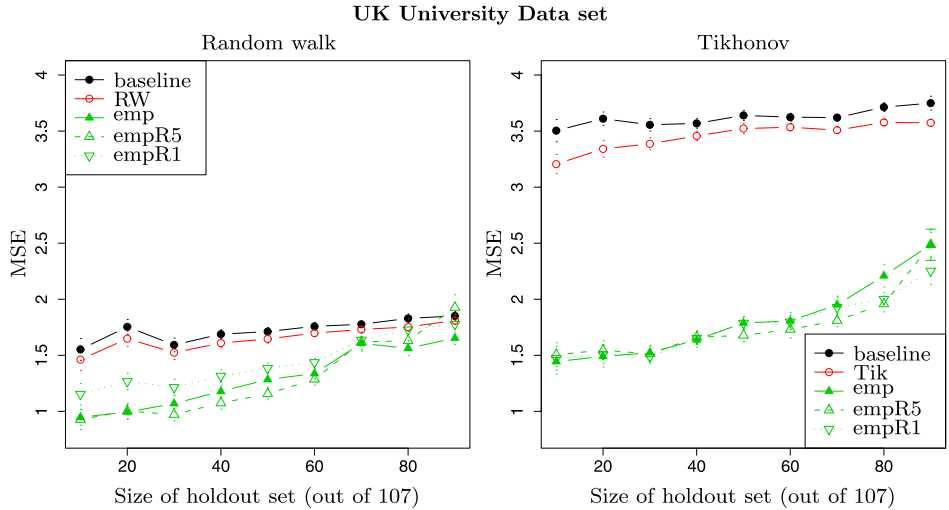}

\caption{MSEs for the RAE scores at different holdout sizes. Left: the
original random walk (red) compared with our empirical random walk
(green). Right: the original Tikhonov (red) compared with our empirical
Tikhonov (green). Baseline methods (black) are described in the text.}
\label{uniMSE}
\end{figure}

The results are shown in Figure \ref{uniMSE}.
The random walk
method performs quite well compared to the Tikhonov method, but
neither of them outperform their corresponding baseline methods by much,
even with the \textit{best} tuning parameters. The black and red curves
track each other closely over a wide range of data holdout sizes,
with the red (graph-based) curve just slightly lower than the
black (baseline) curve.

The results show that the random walk choices $v=X=\sqrt{\pi}$
and $s_{ij} =\pi_iP_{ij}+\pi_jP_{ji}$
are clearly better than the Tikhonov choices
$v=X={\mathbf{1}}_n$ and $s_{ij} = w_{ij}+w_{ji}$ for the UK
university data.
Another difference between the methods is
that the Tikhonov method symmetrizes the graph. As such, it does not distinguish
between links from University $i$ to $j$ and links in
the other direction. Even the baseline for the random walk method,
which does regression on $\sqrt{\pi}$, makes use of the directionality
because that directionality is reflected within $\pi$.

The green curves in Figure \ref{uniMSE} show the error rates for the
two versions of the empirical stationary correlation method. They
generally bring large performance improvements, except at the very
highest holdout levels for the Tikhonov case. Then as few as $17$
University scores are being used and while this is probably too few to
estimate a good covariance, it does not do much harm either. All the
methods do better when less data are held out. The methods with data
driven correlations have slightly steeper performance curves.

We make a numerical summary of the curves from Figure \ref{uniMSE} in
Table \ref{tab:uniResults}. We compare performance for the setting
where about half of the data are held out. For both cases, kriging with
empirical stationary correlations typically brings quite large
improvements over the original methods.
Low rank variations of empirical stationary correlation
kriging perform similarly to the full rank empirical method, except for the
rank $1$ case in the random walk setting. There we still see a large improvement
but not as much as for the full rank or rank $5$ cases.
The good performance of the low rank versions could reflect a small
number of latent effects, or the benefits of regularization.


\begin{table}[t]
\tablewidth=210pt \caption{The relative improvement over baseline when
$50$ of $107$ ARE scores are held out. The baseline methods are simple
regressions through the origin on $X=\sqrt{\pi}$ (random walk) and on
$X={\mathbf{1}}_n$ (Tikhonov)} \label{tab:uniResults}
\begin{tabular*}{210pt}{@{\extracolsep{4in minus 4in}}lcc@{}}
\hline
& \multicolumn{2}{c@{}}{\textbf{Improvement over baseline}}
\\[-6pt]
& \multicolumn{2}{c@{}}{\hrulefill}
\\
 & \textbf{Random walk} & \textbf{Tikhonov}
\\
\hline
Baseline MSE & \phantom{01}$1.71$\phantom{\%} & \phantom{01}$3.64$\phantom{\%}\\[3pt]
Original random walk & \phantom{0}$3.8\%$ & -- \\
Original Tikhonov & -- & \phantom{0}$3.2\%$\\
Empirical &$25.0\%$ & $50.9\%$\\
Empirical R5 & $32.4\%$ & $53.9\%$\\
Empirical R1 & $19.1\%$ & $50.9\%$\\
\hline
\end{tabular*}
\end{table}

\subsection{The WebKB data set}\label{sec:webkb}
The WebKB data set\footnote{The data are at
\url{http://www.cs.umd.edu/projects/linqs/projects/lbc/index.html}.}
contains web pages collected from computer science departments of
various universities in January 1997. The pages were manually
classified into seven categories: student, faculty, staff, department,
course, project and other. The data set we have is a subset, where the
web pages belonging to the ``other'' class are removed. We will only
use the data for Cornell University, which has 195 web pages and 301
links, after removing the three self loops. We further reduce the web
page labels to be ``student'' (1) and ``nonstudent'' ($-1$). There are
83 student pages in total. The adjacency matrix is unweighted, that is,
$w_{ij}$ is 1 if there is a link from page $i$ to $j$ and $0$
otherwise. Again, the links are directed and, hence, $W$ is asymmetric,
with $99.2\%$ of the $w_{ij}$ being zero.

\begin{figure}[t]

\includegraphics{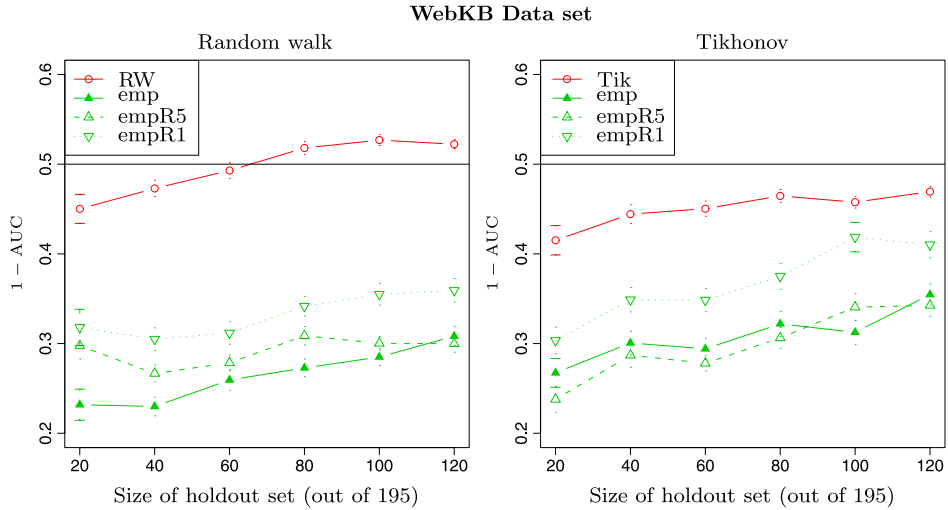}

\caption{Classification error for web page labels at
different holdout sizes, measured with 1 minus the area under the ROC
curve.
Left: the original random walk (red) compared with our empirical random
walk (green). Right: the original Tikhonov (red) compared with our
empirical Tikhonov (green). The baseline method is random guessing.}
\label{webkbAUC}
\end{figure}

\begin{table}[b]
\tablewidth=214pt
 \caption{The relative improvement over baseline when
$100$ out of $195$ web page labels are held out. The baseline AUC is
$0.5$} \label{tab:webkbResults}
\begin{tabular*}{214pt}{@{\extracolsep{4in minus 4in}}lcc@{}}
\hline
& \multicolumn{2}{c@{}}{\textbf{Improvement over baseline}}
\\[-6pt]
&\multicolumn{2}{c@{}}{\hrulefill}
\\
& \textbf{Random walk} & \textbf{Tikhonov}
\\
\hline
Baseline (1 $-$ AUC) & \phantom{0}$0.5$\phantom{\%} & \phantom{0}$0.5$\phantom{\%} \\[3pt]
Original random walk & $-5.4\%$\phantom{.} & -- \\
Original Tikhonov & -- & \phantom{0}$8.5\%$ \\
Empirical & $43.0\%$ & $37.5\%$\\
Empirical R5 & $40.0\%$ & $31.9\%$\\
Empirical R1 & $29.0\%$ & $16.3\%$\\
\hline
\end{tabular*}
\end{table}

The kriging models make continuous predictions of the binary response.
We use the area under the ROC curve (AUC) to measure performance on the
holdout sets. The AUC is equivalent to the probability that a positive
label will get a higher prediction than a negative label. To estimate
the correlation function in the empirical based method, we again use
cubic splines with ten knots for the random walk~$s_{ij}$. However, for
the Tikhonov $s_{ij}$, which has only three possible values 0, 1 and 2
in an unweighted directed graph, we simply use the average at each
$s_{ij}$ without smoothing. The tuning parameters are picked in the
same way as for the university data set.

The results are plotted in Figure \ref{webkbAUC} and summarized in
Table \ref{tab:webkbResults}. As a baseline, we consider a model which
sorts the web pages in random order. It would have an AUC of $0.5$. For
the WebKB data, the Tikhonov method has better accuracy than the random
walk method which actually has trouble getting an AUC below $0.5$. It
is interesting that in this case the method which ignores link
directionality does better. In both cases empirical stationary
correlations bring large improvements. As before, we see that larger
amounts of missing data make for harder prediction problems.

\section{Variations}\label{sec:variation}

In many applications we may want to use more nuanced error variance
measures, such as $\Gamma= \operatorname{diag}(\sigma _1^2,\dots
,\sigma_n^2)$, and this fits easily into the kriging framework. For
example, web pages determined to be spam after a careful examination
could be given a smaller $\sigma^2_i$ than those given less scrutiny,
and those not investigated at all can be given a still higher
$\sigma_i^2$.

Sometimes we can make use of an asymmetry in the labels.
For example, positive determinations, for example,  $1$s, may
have intrinsically higher confidence than negative
determinations, $-1$s, and we can vary $\sigma_i$ to
account for this. Similarly, when one binary label in $\pm1$
is relatively rare, we could use a value other than $0$ as our default
guess.


Finally, it is not necessary to have $v=X$, where $v$ appears in the
variance model through $\sigma^2VRV$ with $V=\operatorname{diag} (v)$
and $X$ in the model for the mean through $\mu X$. We use $v=X$ in the
examples in Section \ref{sec:examples} because this is the choice of
the random walk and the Tikhonov methods. Also, we could hybridize the
Tikhonov and random walk models, using $v=X={\mathbf{1}}_n$ from the
former inside the regression model with the edge directionality
respecting covariance of the latter.


\section{Other related literature}\label{sec:otherlit}

We have so far focused on the graph-based prediction methods from the
machine learning literature. We would like to point out a few related
works in some other fields as well.

In the social network literature, researchers have built network
autocorrelation models to examine social influence process. For more
details see, for example, \citet{Leenders} and
\citet{Marsden}. A typical model is as follows:
%
\begin{eqnarray}\label{eq:autocorrelation}
Y = X\beta+ \omega,\qquad\omega= \alpha B\omega+ \epsilon,
\end{eqnarray}
where $\alpha$ is the network autocorrelation parameter, $B$ is a
weight matrix and $\epsilon\sim\mathcal{N}(0, \sigma^2I)$. This model
is mainly used for estimating or testing $\alpha$ and $\beta$, but we
could of course use it for prediction purpose as well. Notice that we
can write model \eqref{eq:autocorrelation} as
\[
Y\sim\mathcal{N}(X\beta, \sigma^2AA^\prime),
\]
where $A=(I-\alpha B)^{-1}$. Comparing to the other models we have
discussed so far, $Y$ here is no longer a noisy measurement of some
underlying quantity $Z$. The covariance $\sigma^2AA^\prime$ depends on
a scaled weight matrix $\alpha B$. \citet{Leenders} discusses a
few ways to construct the weight matrix $B$, but all of them involve
only the graph adjacency matrix and some a priori quantities.
Nevertheless, the autocorrelation scale $\alpha$, which is estimated
from data, can incorporate some empirical dependence from the observed
$Y$.

\citet{Heaton} consider prediction at unobserved sites in $\mathbb
{R}^1$ or~$\mathbb{R}^2$. The underlying function $Z$ is assumed to
have a sparse wavelet expansion, which they utilize within an MCMC
framework to generate a posterior distribution for the unobserved $Y$.
Their method is shown to have better performance in neighborhoods
containing discontinuities where other methods, for example, kriging,
would smooth. While this method applies to data in Euclidean space with
the regular wavelet transform, \citet{Jansen} discuss a potential
extension to data arising on graphs using the wavelet-like transform
they introduce.

Finally, \citet{Hoff} model the relational tie between a pair of
nodes in a social network by introducing a latent position for each
node in a low dimensional Euclidean space. \citet{Handcock} then
propose a(n) (unsupervised) clustering method by assuming these latent
positions arise from a mixture of distributions, each corresponding to
a cluster. Of course, we can also see the potential to utilize these
latent positions in Euclidean space kriging methods to make
predictions.

\section{Conclusion}\label{sec:conc}

We have shown that several recently developed semi-super\-vised
learning methods for data on graphs can be expressed in terms of
kriging. Those kriging models use implied correlations that derive from
the graph structure but do not take account of sample correlations
among the observed values.

Our proposed empirical stationary correlation model uses
correlation patterns seen among the observed values to estimate a covariance
matrix over the entire graph.
In two numerical examples we saw that using empirical correlations
brought large improvements in performance. Even when there were
large differences between the performance levels of different
semi-supervised methods, the use of empirical correlations
narrowed the gap. This reduces the penalty for the user who
makes a suboptimal choice for $X$, $v$ and $s_{ij}$.

The stationary correlation model was motivated by the
idea that the correlations should be some unknown monotone
function of similarity, and that given enough data,
we could approximate that function.
We were mildly surprised to see a nonmonotone relationship
emerge in our first example, though it was interpretable with hindsight.
We do not have a way to test models of this kind, beyond
using cross-validation to choose between two of them.

We have not implemented our method on any large scale problems. Large
scale presents two challenges. First, solving equations with an
$n\times n$ matrix is expensive. Second, the number of correlation
pairs $\hat R_{ij}$ to smooth is large. Reduced rank correlation
matrices will mitigate the first problem. We might further benefit from
the sparsity of $s_{ij}$ by writing the covariance $\widehat\Psi$ as
sum of a sparse matrix and a rank one matrix. The second problem only
arises when the number $r$ of labeled cases is large. Large $r$ is much
rarer than large $n$, and in any case can be mitigated by downsampling
the correlation pairs before smoothing. In our examples covariance
estimates derived from quite small numbers of observation pairs still
performed well. We finish by pointing out that there are a good many
smaller data sets to which semi-supervised learning on graphs may be
applied.

\section*{Acknowledgments}

We thank the editor, the associate editor and the anonymous referee for
constructive
comments and suggestions that helped improve the paper.

%

\printaddresses

\end{document}